\documentclass[aps,physrev,reprint,superscriptaddress,amsmath,amssymb,floatfix]{revtex4-2}
\usepackage[T1]{fontenc}
\usepackage{lmodern}
\usepackage{graphicx}
\usepackage{xcolor}
\PassOptionsToPackage{hyphens}{url}
\usepackage[colorlinks=true,linkcolor=blue!50!black,citecolor=blue!50!black,urlcolor=blue!50!black]{hyperref}

\makeatletter\g@addto@macro{\UrlBreaks}{\do\_\do\.\do\-}\makeatother

\IfFileExists{YCY_v1/Fig1_layers.pdf}
  {\newcommand{\ycypath}{YCY_v1/}}
  {\newcommand{\ycypath}{}}

\begin{document}

\title{An AI scientist for autonomous quantum microscopy of magnetic minerals}

\author{Changyu Yao}
\thanks{These authors contributed equally.}
\affiliation{Department of Physics, Washington University in St.~Louis, St.~Louis, Missouri 63130, USA}
\author{Yizhou Wang}
\thanks{These authors contributed equally.}
\affiliation{Department of Physics, Washington University in St.~Louis, St.~Louis, Missouri 63130, USA}
\author{Emily G. Wright}
\affiliation{Department of Earth, Environmental, and Planetary Sciences, Washington University in St.~Louis, St.~Louis, Missouri 63130, USA}
\author{Xuanying Lai}
\affiliation{Department of Physics, Washington University in St.~Louis, St.~Louis, Missouri 63130, USA}
\author{Jiahao Yan}
\affiliation{Department of Physics, Washington University in St.~Louis, St.~Louis, Missouri 63130, USA}
\author{Chunhao~Bao}
\affiliation{Department of Physics, Washington University in St.~Louis, St.~Louis, Missouri 63130, USA}
\author{Jeffrey G. Catalano}
\affiliation{Department of Earth, Environmental, and Planetary Sciences, Washington University in St.~Louis, St.~Louis, Missouri 63130, USA}
\author{Chuanwei Zhang}
\email{chuanwei.zhang@wustl.edu}
\affiliation{Department of Physics, Washington University in St.~Louis, St.~Louis, Missouri 63130, USA}
\affiliation{Center for Quantum Leaps, Washington University in St.~Louis, St.~Louis, Missouri 63130, USA}
\author{Chong Zu}
\email{zu@wustl.edu}
\affiliation{Department of Physics, Washington University in St.~Louis, St.~Louis, Missouri 63130, USA}
\affiliation{Center for Quantum Leaps, Washington University in St.~Louis, St.~Louis, Missouri 63130, USA}

\begin{abstract}
Integrating AI agents with scientific instruments creates opportunities for autonomous experimental discovery.
Here we develop an end-to-end AI scientist that operates a quantum diamond microscope to investigate magnetic minerals and connects its measurements with mineralogical interpretation.
With researchers setting goals and reviewing plans, the agent learns to operate the microscope from instrument documentation, calibrates the system, plans and executes measurements, and analyzes the results to decide subsequent experiments.
We apply this workflow to mine tailings from a rare-earth-bearing deposit.
By correlating the quantum magnetic imaging with electron microscopy, the AI scientist identifies nearly ten-fold variation in remanent magnetic moment among individual grains with similar electron image contrast.
These results reveal magnetic heterogeneity that conventional electron microscopy alone does not distinguish, providing complementary information for critical mineral characterization.
By coupling autonomous experimentation with physical reasoning, AI scientists offer researchers a route to scientific discovery beyond their existing expertise while substantially reducing hands-on effort.

\end{abstract}
\maketitle

\section{Introduction}
Magnetic minerals preserve records of geological processes, and their contrasting magnetic properties enable the separation of valuable components from ores and mine tailings~\cite{Glenn2017Geological,Zhang2015LeanIronOres,Li2010TailingsMagneticSeparation,fu2014solar}.
Understanding these properties requires identifying which grains carry magnetism within heterogeneous mineral mixtures. 
Quantum sensing with nitrogen-vacancy (NV) centers in diamond offers a new route to spatially resolve magnetic fields at micrometer scales under ambient conditions~\cite{Maertz2010Vector,Glenn2017Geological,Levine2019QDM,Hsieh2019Imaging,Fu2020Moment,Du2025Geological,Yao2026}, revealing magnetic heterogeneity that may not be apparent from conventional spectroscopy methods.

Translating these magnetic maps into mineralogical insight requires expertise spanning quantum sensing, magnetic reconstruction and mineralogy. 
Such investigations also require adaptive experimentation, in which initial observations constrain physical explanations and guide further measurements to be performed~\cite{Bonato2016Adaptive,Noack2019Scattering}.
Artificial intelligence (AI) agents that combine scientific reasoning with laboratory tools offer an opportunity to integrate autonomous experimental planning, execution and interpretation~\cite{Burger2020Robotic,Boiko2023Autonomous,Bran2024ChemCrow,Sha2025Copilot,Rha2026QUANTA,Bauer2026Transport,Isogawa2026Agentic,Mansouri2026Collaborative,Yan2026QORA,Reed2026Principles,Leutner2026Kerr,Qiu2026AIMS}, helping researchers operate specialized instruments and pursue scientific questions beyond their primary disciplinary expertise.

In this work, we develop an AI scientist for autonomous quantum microscopy of magnetic minerals. 
With researchers setting goals and reviewing plans, the agents learn to operate a commercial quantum diamond microscope from instrument documentation, calibrate the system, plan and execute measurements, and analyze the results to guide further experiments. 
We apply this workflow to iron-ore tailings from the rare-earth-bearing Pea Ridge deposit~\cite{Day2016StFrancois, Harlov2016PeaRidge} and identify predominantly remanent magnetism.
By correlating magnetic maps with electron microscopy images, the AI scientist reveals nearly ten-fold variation in reconstructed magnetic moment among grains with similar electron-image contrast, and helps distinguish magnetite from hematite at the single-grain level.
Together, these results demonstrate how AI agents can connect autonomous quantum experimentation with mineralogical interpretation for magnetic mineral characterization.

\begin{figure*}[t]
  \centering
  \includegraphics[width=\textwidth]{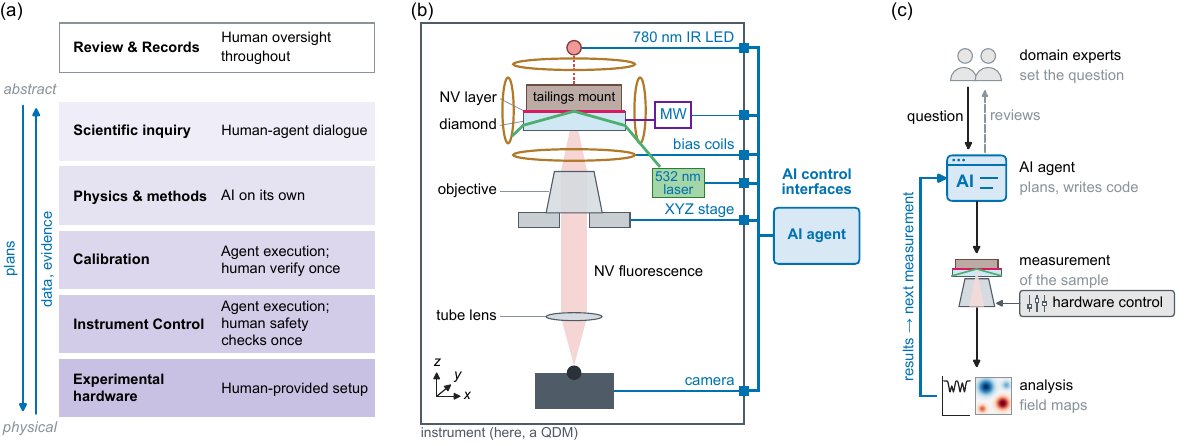}
  \caption{An agent-based workflow for experimental research. (a) Roles of researchers and the agents at each level, from the physical instrument to scientific interpretation. (b) The agent uses instrument manuals and online documentation to build control tools, then uses those tools to automate measurements on the target instrument, here a commercial quantum diamond microscope (QDM). (c) Closed-loop measurement under expert guidance. Researchers set the scientific direction. The agent can run the full loop by planning and executing each measurement, analyzing the data, and choosing the next measurement from the result.}%
  \label{fig:ycy-1}
\end{figure*}

\section{An AI scientist operating a quantum diamond microscope}

Our AI scientist connects experimental planning, instrument operation and physical interpretation in an iterative workflow (Fig.~\ref{fig:ycy-1}a,c). 
Researchers set the scientific goals and verify the initial control and calibration procedures. 
The agents translate these goals into measurement plans, execute experiments and analyze the results to guide subsequent measurements. 
Within the agreed plan, routine operations proceed without step-by-step confirmation, while researchers review the emerging scientific interpretation and can intervene throughout. 
The agents automatically record all measurement plans, hardware operations and analysis outputs to support this review.

We implement this workflow on a commercial widefield quantum diamond microscope from Quantum Catalyzer~(Fig.~\ref{fig:ycy-1}b). 
The sample is positioned close to a thin layer of NV centers near the diamond surface. 
A 532~nm laser excites the NV centers, a microwave source drives their spin transitions, and the resulting fluorescence is imaged onto a camera. 
Three pairs of Helmholtz coils provide the bias magnetic field, while a 780~nm LED supports infrared sample imaging and a motorized XYZ translation stage enables positioning and focusing. 
Starting from instrument documentation, the agents develop control tools that coordinate these six hardware interfaces to acquire images and quantum sensing spectra, reconstruct magnetic field maps and execute subsequent measurements. 
Researchers prepare the samples, manage physical changes to the apparatus and retain responsibility for experiment safety.

\begin{figure*}[t]
  \centering
  \includegraphics[width=\textwidth]{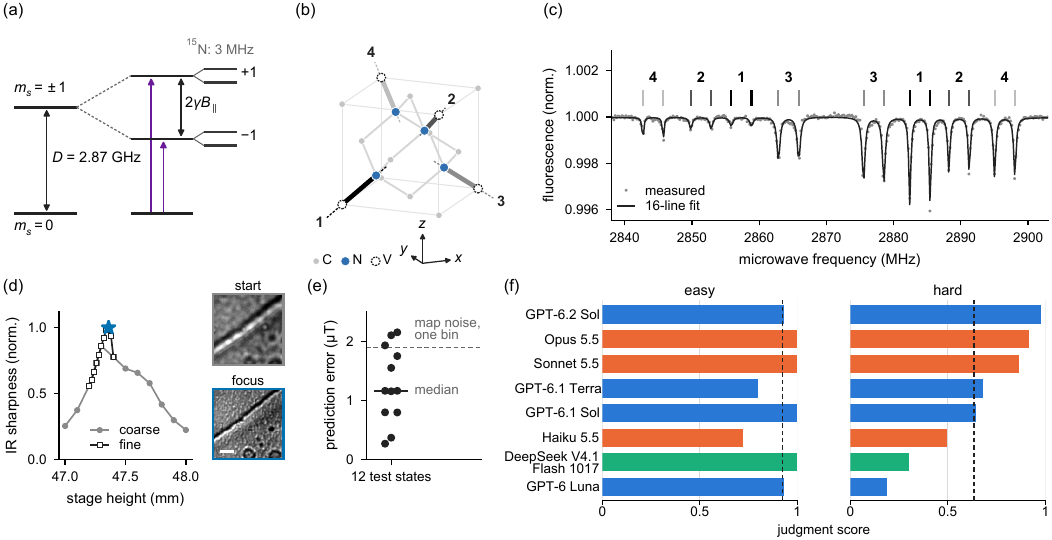}
  \caption{(a) Ground-state spin levels of NV center in diamond with $^{15}$N nuclear isotope. The zero-field splitting is $D=2.87$ GHz. A field component $B_{\parallel}$ along the NV axis separates the $m_s=\pm1$ levels by $2\gamma_e|B_{\parallel}|$, where $\gamma_e\simeq28$ GHz/T. Coupling to the $^{15}$N nuclear spin ($I=1/2$) splits each transition into two lines about 3 MHz apart. (b) The four NV orientations in the diamond lattice. (c) Experimental ODMR spectrum at a bias field of 1.10 mT, with a 4-pairs of resonances fit. Each numbered NV orientation contributes two transitions, each with a $^{15}$N hyperfine doublet. (d) Coarse and fine scans of infrared image sharpness versus stage height. Sharpness is the mean squared gradient of the smoothed image divided by the squared mean intensity. The star marks the agent selected focus. The insets show the infrared image before and after focusing. (e) Magnitude of the difference between the field vector predicted from the coil currents and measured by the NV. The median prediction error was $1.2\,\mu$T for 12 different test fields. The dashed line marks the photon shot noise level of the NV magnetometer at each image bin. (f) Scores of eight AI models on five calibration tasks. The dashed lines mark the mean score across models for each task group.}
  \label{fig:ycy-2}
\end{figure*}

The NV center is a point defect formed by a substitutional nitrogen atom adjacent to a lattice vacancy~\cite{Doherty2013PhysRep}.
The defect hosts a spin-triplet ground state ($S=1$) with a zero-field splitting (ZFS) of $D \approx 2.87$~GHz between the $|m_s=0\rangle$ and $|m_s=\pm1\rangle$ sublevels (Fig.~\ref{fig:ycy-2}a). 
Under 532~nm excitation, a spin-dependent intersystem crossing enables efficient optical polarization into the $|m_s=0\rangle$ state and readout through spin-dependent photoluminescence, where $|m_s=0\rangle$ emits more photons on average than $|m_s=\pm1\rangle$.
The spin transitions can be measured using optically detected magnetic resonance (ODMR): by sweeping the frequency of an applied microwave field while monitoring the fluorescence intensity, a drop in fluorescence is observed when the microwave is resonant with the $|m_s=0\rangle \leftrightarrow |m_s=\pm1\rangle$ transitions. 

An external magnetic field lifts the degeneracy of the $|m_s=\pm1\rangle$ states via the Zeeman effect, with splitting proportional to the field projection along the NV axis.
In our experiment, single-crystal diamond contains a dense ensemble of NV centers with four distinct NV orientations (Fig.~\ref{fig:ycy-2}b).
These orientations give rise to up to four pairs of ODMR resonances, encoding four projections of the local magnetic field and enabling reconstruction of its vector components~\cite{Levine2019QDM,Wang2025Nanodiamond} (Fig.~\ref{fig:ycy-2}c).

To validate the AI scientist’s control of the quantum diamond microscope, we first task it with autonomous calibration of the bias field generated by three pairs of Helmholtz coils. 
The agents begin by adjusting the microscope focus to maximize the sharpness of the infrared camera image (Fig.~\ref{fig:ycy-2}d). 
They then vary the coil currents $I_k$ ($k=x,y,z$ in the lab frame), acquire ODMR spectra and fit the resonances to extract the magnetic field, $\vec{B}$, at the NV layer.
These measurements determine the linear relation, $\vec{B}=\vec{B}_0+\sum_k I_k\vec{c}_k$, where $\vec{B}_0$ is the ambient field (i.e., Earth's magnetic field) and $\vec{c}_k$ is the vector field generated per unit current by each coil pair accounting for the imperfect alignment in the experiment.

The calibration must also resolve the sign ambiguity between \(\vec{B}\) and \(-\vec{B}\) and establish the transformation from the NV crystal frame to the laboratory frame. 
The agents devise procedures that use ambient field $\vec{B}_0$~\cite{Beggan2026IGRF14} to determine the sign and a consistency check based on Maxwell’s equations~\cite{Levine2019QDM,Yao2025Universal} to determine the in-plane rotation. 

We validate the calibration at 12 additional current settings excluded from fitting (Fig.~\ref{fig:ycy-2}e). The median difference between predicted and measured field vectors is \(1.2\,\mu\mathrm{T}\), comparable to the \(1.9\,\mu\mathrm{T}\) photon shot noise level of an individual magnetic-image bin. 
The calibrated coil strengths and ambient field also agree closely with an earlier human calibration of the same instrument~\cite{Wang2025Nanodiamond}.

Using the saved calibration data, we benchmark eight AI models on five calibration tasks~(Fig.~\ref{fig:ycy-2}f)~\cite{Kapoor2024AgentsThatMatter,Chen2025ScienceAgentBench,Cai2026Instruments,Gomez2026HERA}.
Most models perform well on easier tasks including microscope focus and coil polarity, whereas harder tasks, such as extracting precise coil-response vectors $\vec{c}_k$, are more consistently solved by the strongest models (GPT-6.2 Sol, Opus 5.5 and Sonnet 5.5).
These results suggest that routine calibration analysis can be delegated to smaller models, while demanding tasks benefit from more capable models.

\begin{figure}[tbp]
  \centering
  \includegraphics[width=\columnwidth]{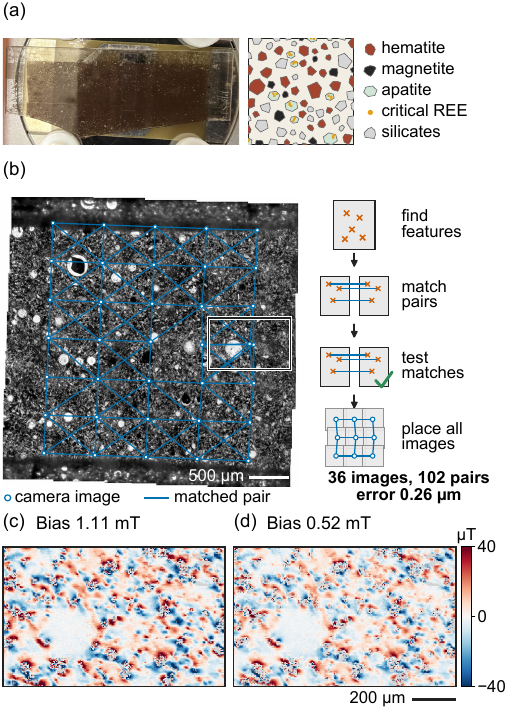}
  \caption{(a) Polished thin section of mine tailings from the Pea Ridge iron oxide--apatite deposit in Missouri. The schematic illustrates minerals found in the tailings, including hematite, magnetite, and fluorapatite, a host for rare-earth elements; critical REE-bearing phosphates (monazite and xenotime) occur both as inclusions in apatite (shown) and as discrete grains. (b) Transmitted infrared mosaic acquired on the QDM. The agents stitch 36 images by matching features in overlapping tiles. Circles mark tile centers, lines connect matched pairs, and the box marks the field for magnetic imaging in (c,d). (c,d) Sample stray field projected along the bias field direction at strength 1.11 and 0.52 mT, respectively. 
  }
  \label{fig:ycy-3}
\end{figure}

\section{Autonomous investigation of magnetic heterogeneity in minerals}

We next apply the AI scientist to investigate a polished thin section of iron-ore tailings from the rare-earth-bearing Pea Ridge deposit in Missouri~\cite{Day2016StFrancois,McCafferty2016Magnetic} (Fig.~\ref{fig:ycy-3}a). 
Its grains contain mainly magnetic magnetite (ferrimagnetic) and hematite (weakly magnetic) 
, as well as non-magnetic apatite, magnesium silicate, quartz, and minor rare-earth (REE) phosphates (monazite and xenotime)~\cite{Day2016StFrancois,McCafferty2016Magnetic,Aleinikoff2016PeaRidge,Ayuso2016PeaRidge,Harlov2016PeaRidge}.
All quantum diamond microscopy measurements and associated analyses are performed by the agents, while researchers provide scientific direction and periodically review the plans and interpretations. 
Between reviews, the agents execute measurements, analyze the results and guide further investigation without step-by-step human control. 

The quantum diamond microscope has a typical field of view of approximately $1\times0.64$~mm$^2$, covering only a small portion of the sample in each acquisition. 
To establish a spatial reference over a larger area, the agents move the sample stage, acquire overlapping infrared images and align them using shared image features. 
They assemble 36 images into a mosaic spanning \(3.7\times3.6\)~mm$^2$ (Fig.~\ref{fig:ycy-3}b), which provides a common coordinate system for locating grains and registering subsequent magnetic measurements.
They then acquire spatially resolved ODMR spectra across the surveyed area and reconstruct maps of the sample’s stray magnetic field.
Fields of a few tens of microtesla are observed near many of the grains (Fig.~\ref{fig:ycy-3}c).

To distinguish remanent magnetism from a response induced by the bias magnetic field, the agents compare magnetic maps acquired at different bias-field strengths.
Reducing the bias from \(1.11\) to \(0.52\)~mT decreases the stray-field amplitude by only about \(11\%\), substantially less than expected for a purely linear induced response (Fig.~\ref{fig:ycy-3}d).
Reversing the bias field also preserves the sign of the dominant stray-field component.
Together, these measurements establish predominantly remanent magnetism, motivating reconstruction of the magnetic moments carried by individual grains.

\begin{figure*}[t]
  \centering
  \includegraphics[width=\textwidth]{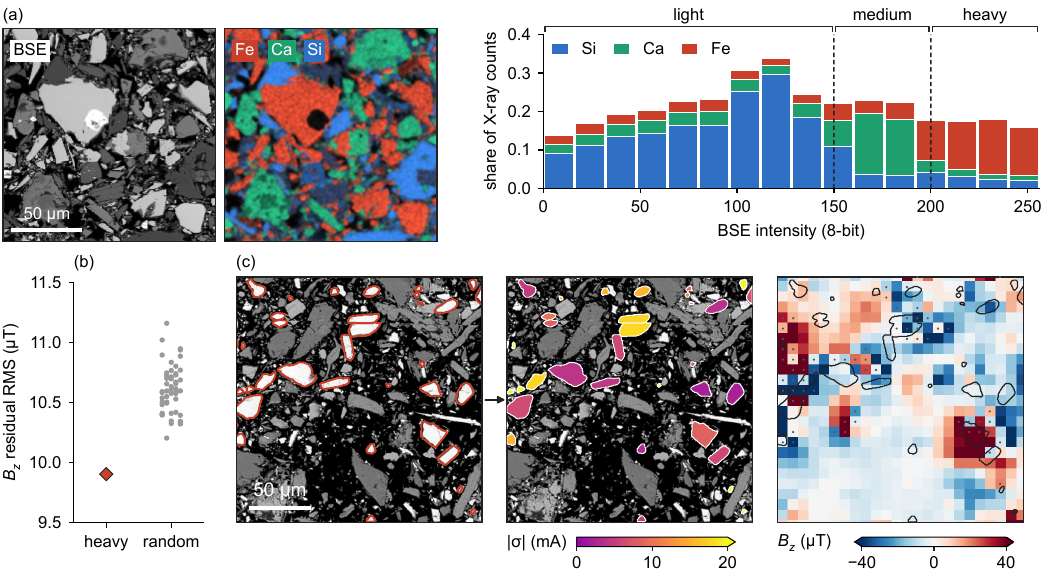}
  \caption{(a) Left: Backscattered-electron (BSE) image and energy-dispersive X-ray spectroscopy (EDS) map of the same reference region. Right: the correlated BSE intensity with X-ray counts. The dashed lines mark the boundaries between the light, medium, and heavy BSE intensity classes. Fe-rich material is most abundant in the brightest BSE class (heavy grains). (b) Grain-level magnetization is reconstructed following Ref.~\cite{Yao2025Universal} with candidate sources restricted either to only the heavy grains (red diamond) or to 50 random grain sets matched in area, size distribution, and degrees of freedom (gray dots). The heavy-grain reconstruction yields a lower root-mean-square (RMS) residual field than every randomized control. This agrees with the concentration of iron oxides in the heavy grains. (c) Heavy grains outlined on the BSE image (left), colored by their reconstructed remanent moment per unit area, $|\sigma|$ (middle), and overlaid on the measured out-of-plane stray field $B_z$ at a bias field of $0.52\,\mathrm{mT}$ (right).  Dots in the field map mark bins with broadened NV resonances, consistent with strong local stray-field gradients near grains with large magnetic moments per unit area, \(|\sigma|\)~\cite{Hsieh2019Imaging,Levine2019QDM}.}
  \label{fig:ycy-4}
\end{figure*}

To connect the magnetic response with mineralogy, the agents analyze existing scanning electron microscopy (SEM) images and energy-dispersive X-ray spectroscopy (EDS) data. 
In the backscattered-electron (BSE) images, brighter contrast generally indicates a higher average atomic number, while EDS provides elemental composition. 
Measurements in reference regions on the same sample mount associate the brighter grains, termed heavy grains with BSE intensity $\ge 200$, with Fe-rich material (Fig.~\ref{fig:ycy-4}a). %
Co-localized O signal indicates that this population consists predominantly of iron oxides, including ferrimagnetic magnetite (\(\mathrm{Fe_3O_4}\)) and weakly magnetic hematite (\(\alpha\text{-}\mathrm{Fe_2O_3}\)), which the available electron images and elemental maps do not uniquely distinguish. 
Their contrasting magnetic properties provide an opportunity for quantum magnetic imaging to help distinguish these iron-oxide phases.

The agents register the magnetic map to its SEM image. 
Identifying the magnetic contribution of each grain requires reconstructing its source moment from the overlapping stray fields of neighboring grains.
Guided by the compositional evidence, the agents restrict the candidate magnetic sources to only the heavy-grain population and fit an effective remanent magnetic moment per unit area, $\sigma$, for each grain~\cite{Fu2020Moment,Yao2025Universal}. 
The magnetic field predicted from the fitted heavy-grain moments closely matches the measured field, with a smaller residual than those obtained using randomly selected sets of source grains (Fig.~\ref{fig:ycy-4}b). This comparison further supports restricting the magnetic sources to the heavy-grain population.

Among 70 grains large enough to resolve, the reconstructed remanent areal moment density, $|\sigma|$, varies nearly ten-fold despite similar electron-image contrast (Fig.~\ref{fig:ycy-4}c). 
The strongest grains exceed the modeled hematite limit~\cite{Dunlop2001Hematite}, supporting the presence of magnetite-like carriers. 
Most remaining heavy grains exhibit both modest remanence and weak induced response, consistent with a predominantly hematite population, although fine, high-coercivity magnetite cannot be excluded~\cite{Dunlop1997Rock}.
Together, these results demonstrate how an AI scientist combines autonomous quantum diamond microscopy with mineralogical reasoning to help distinguish magnetic mineral phases at single-grain resolution.

\section{Outlook}
We demonstrate an AI scientist that interfaces with a commercial quantum diamond microscope and connects autonomous measurements with mineralogical reasoning to resolve magnetic heterogeneity at the single-grain level. 
Looking forward, our work points to immediate opportunities for strengthening agentic experimentation.
While agents already automate hardware control and routine analysis in our workflow, human oversight remains essential for selecting decisive measurements, assessing the confidence of scientific interpretations, and enforcing experimental safety limits.~\cite{Ruan2024ToolEmu,Isogawa2026Agentic,Reed2026Principles}.
Integrating richer physical models, digital twins of laboratory instruments, and targeted validation experiments into agent workflows offers a practical near-term path to detecting failures missed by routine checks and expanding the scope of reliable autonomous experimentation.

More broadly, the same workflow could support diamond quantum sensing in materials science and biology, from magnetic order and superconductivity~\cite{Casola2018Probing,liu2025quantum} to local thermal and magnetic response in living systems~\cite{Schirhagl2014Nitrogen,kashem2026multiplexed}.
Learning instrument operation from documentation and real experiments offers a path to extending this approach to other quantum measurement platforms and classical laboratory instruments.
Operating across multiple instruments could in turn allow AI scientists to coordinate quantum measurements with complementary imaging and spectroscopy to test competing physical explanations and guide further experiments.
With researchers setting goals and verifying key conclusions, such systems can substantially reduce hands-on experimental effort and enable investigations beyond an individual researcher’s primary expertise, pointing toward AI scientists that learn increasingly general capabilities through direct interaction with the physical world.

\begin{acknowledgments}
We thank James Kennedy of Caldera Holding LLC for providing access to the tailings at the Pea Ridge mine. We thank Jeffrey Gillis and Jennifer Houghton for access to and assistance with SEM--EDS, Christopher Jones of Halliburton for valuable discussion. Y.W. thanks Haochen Shen for discussion, and Alex Chen for compute resources. 
C.Y., C.Zhang and C.Zu acknowledge support from the U.S. Department of Energy, Office of Science, Basic Energy Sciences, under Award Number DE-SC0026580 for developing the AI scientists.
X.L. acknowledges support from the Edwin Jaynes Fellowship in the Department of Physics at Washington University in St. Louis.
J.G.C. and E.G.W. were supported by the National Science Foundation Directorate for Geosciences under Award No.~2327617. E.G.W. was also supported by a National Science Foundation Graduate Research Fellowship under Award No.~2139839.
C. Zhang acknowledges support from NSF grants 2503230 and 2607484. C. Zu acknowledges support from NSF 2514391 and NSF NRT LinQ 2152221.
We also acknowledge institutional support from the Center for Quantum Leaps at Washington University.
\end{acknowledgments}

\bibliography{\ycypath refs_ycy_v1}
\end{document}